\documentclass{article}

\PassOptionsToPackage{numbers, sort&compress}{natbib}
\usepackage[main, final]{neurips_2026}

\usepackage[utf8]{inputenc} 
\usepackage[T1]{fontenc}    
\usepackage{hyperref}       
\usepackage{url}            
\usepackage{booktabs}       
\usepackage{amsfonts}       
\usepackage{nicefrac}       
\usepackage{microtype}      
\usepackage{xcolor}         
\usepackage{graphicx}
\usepackage{amsmath, amssymb}
\usepackage{subcaption}
\usepackage{multirow}

\newcommand\todo[1]{\textcolor{black}{#1}}
\newcommand\toSonia[1]{\textcolor{black}{#1}}
\newcommand\HZ[1]{\textcolor{black}{#1}}

\title{Qubrio: High-Performance Quantum Compilation via Multi-Agent LLM Collaboration}

\author{%
  Jixuan Ruan$^{1*}$ \quad Zhuo Cui$^{1*}$ \quad Zhengding Hu$^{1}$ \quad Zhongkai Yu$^{1}$ \quad Xiang Fang$^{1}$ \\
  \textbf{Yue Guan$^{1}$ \quad Jason Ludmir$^{2}$ \quad Phillip E. Weinberg$^{3}$ \quad Xiu-Zhe Luo$^{3}$} \\
  \textbf{Sheng-Tao Wang$^{3}$ \quad Sonia Lopez Alarcon$^{4}$ \quad Hezi Zhang$^{5}$ \quad Yufei Ding$^{1}$} \\
  $^{1}$University of California, San Diego \quad $^{2}$Rice University \quad
  $^{3}$QuEra Computing Inc. \\  $^{4}$Rochester Institute of Technology \quad $^{5}$University of Wisconsin–Madison
}

\begin{document}

\maketitle
{\renewcommand\thefootnote{}\footnotetext{$^*$Equal contribution.}}

\begin{abstract}
Large Language Models (LLMs) offer an adaptable alternative to traditional quantum compilation heuristics, which bottleneck progress by demanding costly manual redesigns whenever hardware evolves. However, naive single-pass LLM approaches fail under \todo{the unique challenges introduced by quantum compilation.}
We introduce Qubrio, a practical agentic framework that significantly reduces heuristic redesigns through a three-level decomposition. By (1) partitioning \HZ{programs} into sequential operation stages, (2) deploying specialized agents for placement, routing and optimization, and (3) using disaggregated feedback for precise error correction, our system reliably navigates complex physical constraints. 
\toSonia{Beyond serving as a direct compiler,} our LLM framework also uncovers novel strategies that facilitate convoy-style shuttling, substantially enhancing hardware concurrency without sacrificing fidelity. We further integrate these strategies into the previous state-of-the-art (SOTA) compiler, recovering a significant fraction of the LLM's performance advantages.
On realistic workloads, our compiler achieves significant improvements in hardware runtime and \HZ{program} fidelity over the SOTA baseline PowerMove, seamlessly adapting to new hardware capabilities. Qubrio features an open user interface for practical use, and its code is publicly available at \url{https://github.com/crosaa/Qubrio}.
\end{abstract}

\section{Introduction}

Quantum compilation, the process of translating high-level descriptions of quantum algorithms into hardware-executable instructions, is a key challenge in enabling practical quantum computation \cite{zhu2025quantum, maronese2022quantum, yan2024quantum, ruan2025powermove, bochen2024compilation}.
Existing quantum compilers \cite{tan2024compiling, wang2024atomique, ruan2025powermove, bochen2024compilation} typically rely on  
domain-engineered heuristics to handle diverse hardware constraints and capabilities. This reliance, however, creates a significant maintenance burden. Adapting to new hardware features often requires domain experts to redesign heuristics and rewrite substantial portions of the compilation pipeline \cite{zhu2025quantum, bandic2022full, wang2024atomique, ruan2025powermove}. This highlights a critical need for compilation frameworks capable of automatically adapting to evolving hardware.

Against this backdrop, LLMs \cite{achiam2023gpt, liu2024deepseek} offer a promising solution. Notably, LLMs have demonstrated strong capabilities in program synthesis, code optimization, and reasoning over structured representations \cite{hong2023metagpt, funsearch, shinn2023reflexionlanguageagentsverbal, chen2021evaluating}. This suggests they can learn compilation logic directly from specific hardware descriptions, significantly reducing the need for experts to redesign heuristics. However, whether and how LLMs can work as practical quantum compilers is still largely unexplored.

Empirical observations reveal that naive single-pass LLM compilation fails in practice. We attribute this failure to three orthogonal sources of pressure that any LLM compilation agent must absorb, corresponding to the input, processing, and feedback stages of its operational loop: (i) Input complexity: A full quantum \HZ{program} exceeds the regime in which current LLMs reason reliably. Current LLMs struggle with long contexts, especially in tasks like quantum compilation where each decision affects all the decisions that follow. 
(ii) Task heterogeneity: Compilation couples sub-tasks with distinct reasoning modes (e.g., spatial assignment versus sequential trajectory planning). Forcing a single LLM agent to handle these jointly induces cross-task interference, degrading performance on all of them.
(iii) Feedback ambiguity:
LLMs improve their outputs by iterating on feedback, but feedback that only reports overall performance does not tell the model which constraint was violated and which step caused this violation, leaving the agent without an actionable gradient for repair.

We address the challenges through a three-level decomposition that relieves pressure at each stage of the compilation process. (1) Input decomposition partitions a quantum program into sequential operation stages, so that each stage fits within the context length over which LLMs reason reliably. (2) Task decomposition assigns sub-tasks to specialized agents (placement, routing and optimization) each matched to one reasoning mode, eliminating cross-task interference. (3) Feedback decomposition replaces overall performance scores with structured, multi-dimensional signals: the verifier localizes constraint violations to the offending step, and the simulator disentangles performance into distinct physical error sources, enabling the agent to reason about trade-offs and apply targeted repair.

Furthermore, our LLM framework uncovers novel strategies that facilitate a highly parallel routing pattern, which we term 'convoy-style shuttling'.
Unlike previous compilers that route qubits  along shortest paths, which frequently induces move conflicts under hardware constraints, the LLM learns to leverage buffer positions to resolve these conflicts and align multiple qubits into synchronized movements.
Moreover, the LLM learns to maintain qubits in continuous motion across consecutive routing rounds to minimize associated fidelity penalties introduced by qubit transfer (pick-up and drop) operations.
This coordination parallelizes otherwise sequential movement operations, significantly accelerating hardware execution without compromising fidelity.
A key insight is that without the bias of a hardcoded heuristic algorithm, the LLM uncovers non-trivial strategies to accelerate hardware execution. This demonstrates our framework's dual value: not just as a practical compiler, but as a systematic tool for discovering novel heuristics.



In summary, our contributions are as follows:
\begin{itemize}
    \item[\textbf{(1)}] \textbf{An Adaptable Agentic Compiler Framework:} We introduce an agentic quantum compilation framework that significantly reduces the need for manual heuristic rewrites. To overcome the failures of single-pass LLMs, we propose a three-level decomposition framework: partitioning \HZ{programs} into manageable stages, deploying specialized agents for distinct tasks, and providing disaggregated feedback for precise correction and optimization.

    \item[\textbf{(2)}] \textbf{LLM-Driven Heuristic Discovery and Convoy-Style Shuttling:} \toSonia{Beyond serving as a direct compiler,} our framework also uncovers novel strategies that utilize buffer positions and continuous motion to facilitate convoy-style shuttling, substantially enhancing hardware concurrency without sacrificing fidelity. We further integrate these strategies into the classical compiler, recovering a significant fraction of the LLM's performance advantages.

    \item[\textbf{(3)}] \textbf{SOTA Efficacy and Hardware Adaptability:} Qubrio consistently outperforms the SOTA compiler PowerMove\cite{ruan2025powermove}, yielding average improvements of \todo{4.7$\times$} and \todo{1.3$\times$} in hardware runtime and fidelity, and easily adapts to new hardware capabilities. 
    
\end{itemize}

\section{Background}

\begin{figure*}[tp]
 \centering
  \includegraphics[width=0.9\linewidth, trim=0.2in 0.2in 0.3in 0in, clip]{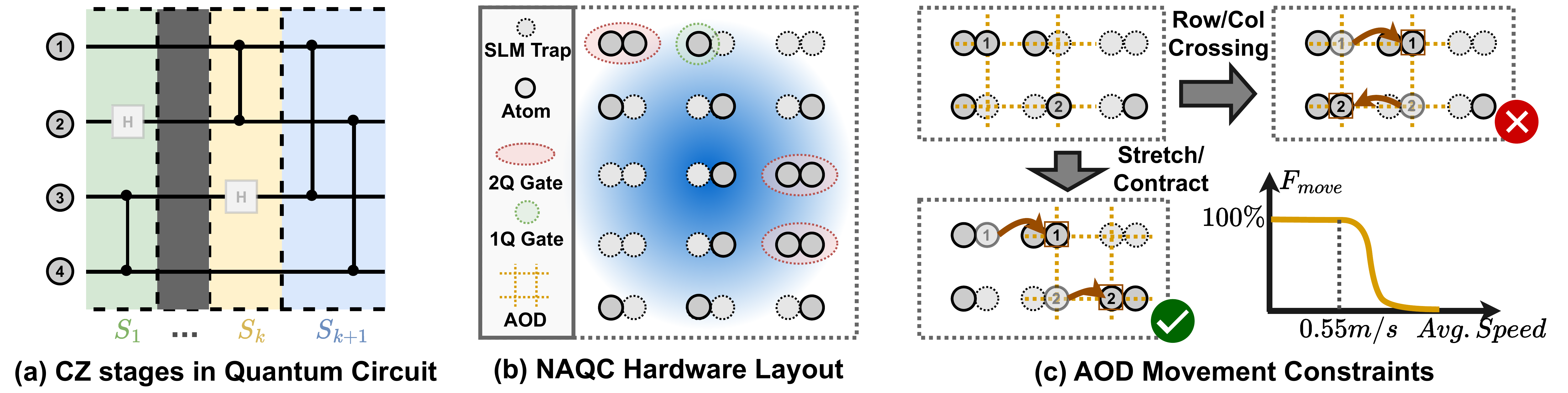}
  \caption{\textbf{Quantum circuit structure and neutral atom hardware.} (a) Circuits decompose into sequential stages of parallel CZ interleaved with single-qubit gates. (b) Atoms reside in static SLM and are shuttled into interaction distance for two-qubit gates. (c) Atom movements prohibit row and column crossings, requiring controlled transport velocities to ensure high movement fidelity ($F_{move}$).}
  \label{fig:background}
\end{figure*}

\noindent\textbf{Quantum Compilation Problem.}
Quantum compilation translates a high-level quantum \HZ{program, typically represented by a quantum circuit,} into executable operations on a  \HZ{quantum processor with hardware constraints} \cite{zhu2025quantum, maronese2022quantum, yan2024quantum, ruan2025powermove, bochen2024compilation}. Because hardware limits two-qubit interactions to adjacent physical positions, a common strategy is to decompose an $n$-qubit circuit into a temporal sequence of stages, $\{S_1, \ldots, S_K\}$ \cite{bochen2024compilation, ruan2025powermove}. Each stage contains parallelizable two-qubit gates \todo{interleaved with} single-qubit gates (Fig.~\ref{fig:background}(a)). To satisfy the hardware constraints, the compiler must solve two coupled optimization tasks per stage: (1) Placement (mapping logical qubits to physical positions) and (2) Routing (moving qubits to adjacent \HZ{positions} for two-qubit gate interaction). Concatenating these local solutions yields a global execution schedule. The goal of the compiler is to generate a valid, executable schedule that minimizes total hardware runtime while maximizing overall fidelity.

\noindent\textbf{Neutral Atom Quantum Computing.} 
Neutral atom quantum computers (NAQCs) have emerged as one of the leading platforms for scalable quantum computation~\cite{ebadi2021quantum, graham2022multi, wurtz2023aquila, bluvstein2024logical, bluvstein2022quantum, bluvstein2025architectural}, offering an attractive combination of long coherence times~\cite{ebadi2021quantum, graham2022multi, wurtz2023aquila, bluvstein2024logical}, high-fidelity operations~\cite{evered2026highfidelityentanglinggatesnonlocal, bluvstein2024logical, levine2022dispersive, bluvstein2022quantum, levine2019parallel, fu2022high, evered2023high}, \HZ{and flexible reconfigurability \cite{bluvstein2022quantum, bluvstein2024logical}}. However, unlocking this potential requires a compiler that can navigate the platform's uniquely demanding routing problem: unlike the localized SWAP operations in superconducting and trapped-ion systems~\cite{zhu2025s, li2019tackling}, NAQC routing requires \HZ{global} movement planning across the entire qubit array. We instantiate our framework on this platform, as success under its stringent constraints  implies broad applicability to simpler routing paradigms.
NAQCs execute computations by trapping individual atoms in a 2D grid of static traps generated by a spatial light modulator (SLM) (Fig.~\ref{fig:background}(b)), and routing them via mobile traps controlled by an acousto-optic deflector (AOD) (Fig.~\ref{fig:background}(c)) \cite{bluvstein2022quantum, bluvstein2024logical}.
While single-qubit gates execute in place, executing two-qubit gates requires transferring target atoms from SLMs into AODs \HZ{and then} shuttling them to adjacent traps for interaction \cite{bluvstein2022quantum,bluvstein2024logical, jaksch2000fast, urban2009observation, levine2019parallel}.
This movement is governed by three critical hardware constraints (Fig.~\ref{fig:background}(c)): (1)~Topological Ordering:
The AOD grid can stretch or contract, but rows and columns can never cross \cite{bluvstein2022quantum, bluvstein2024logical}. 
(2)~Velocity Bounds: Shuttling speed must be controlled to prevent atom loss \cite{bluvstein2022quantum,bluvstein2024logical, bochen2024compilation,ruan2025powermove}.
(3)~Transfer Penalties: Each SLM--AOD transfer incurs a non-negligible error \cite{bluvstein2022quantum, bluvstein2024logical}. Minimizing these transfers often conflicts with minimizing overall routing time, requiring the compiler to carefully navigate this trade-off.

\noindent\textbf{LLM Agents for Structured Reasoning.}
While LLMs and agentic frameworks \cite{chen2021evaluating, austin2021program, wei2022chain} have demonstrated strong capabilities in structured reasoning, their naive application to quantum compilation proves inadequate.
\todo{To reason under strict constraints, general agentic paradigms typically rely on two design patterns:} (1) multi-agent decomposition 
, which allocates subtasks to specialized agents to minimize cross-task interference \cite{chen2023communicative, hong2023metagpt}; and (2) verification-in-the-loop, which couples LLM generation with deterministic checkers to bridge the gap between probabilistic outputs and strict correctness guarantees \cite{chen2023teachinglargelanguagemodels, aggarwal2024alphaverus, shinn2023reflexionlanguageagentsverbal}. 
\todo{However, the unique challenges of quantum compilation render these generic task divisions and standard verify-then-fix loops ineffective, necessitating domain-specific designs.}


\section{Method}\label{sec:method}


Fig. \ref{fig:overview} illustrates the full pipeline, which is explicitly structured around our three-level decomposition strategy. First, to manage input complexity, input decomposition partitions the circuit into manageable CZ stages $\{S_1, \ldots, S_K\}$. Second, to prevent cross-task interference, task decomposition delegates distinct reasoning modes to specialized modules: a Placement Agent (Fig. \ref{fig:overview}(e)) assigns qubits to trap positions, while a Routing Agent (Fig. \ref{fig:overview}(f)) generates AOD movement operations. Finally, addressing feedback ambiguity, feedback decomposition is implemented via dedicated verifiers. The Placement Verifier (Fig. \ref{fig:overview}(b)) and Routing Verifier (Fig. \ref{fig:overview}(c)) isolate and localize specific constraint violations. Once a stage is successfully placed and routed, an Optimize Agent (Fig. \ref{fig:overview}(g)) refines the local schedule through a generate–evaluate loop guided by a Hardware Simulator (Fig. \ref{fig:overview}(d)).


\subsection{Placement Agent}
The Placement Agent (Fig. \ref{fig:overview}(e)) computes the qubit-to-trap mapping $P_{k+1}: Q_{k+1} \rightarrow T$ for each CZ stage $S_{k+1}$. Given the hardware specification and the previous mapping $P_k$, the agent optimizes $P_{k+1}$ to ensure all CZ pairs stay within interaction distance $r$ while maintaining a separation $>r$ for non-interacting pairs to eliminate crosstalk.

\textbf{Hierarchical verification and feedback.} To resolve mapping violations without causing optimization oscillations, the verifier localizes errors into three structured sets of tuples $(q_i, q_j, P_{k+1}(q_i), P_{k+1}(q_j))$, providing the model with entity-level localization:
\begin{itemize}
    \item Collision ($V_{coll}$): Qubits assigned to the same trap, where $P_{k+1}(q_i) = P_{k+1}(q_j)$.
    \item Adjacency ($V_{adj}$): CZ pairs $(q_i, q_j) \in S_{k+1}$ erroneously placed at distance $> r$.
    \item Crosstalk ($V_{xtalk}$): Non-interacting pairs erroneously placed at distance $\leq r$.
\end{itemize}

Constraints are addressed via a two-stage reasoning pipeline. (1) Stage 1 prioritizes $V_{coll}$ independently to establish a valid topological baseline. Once $V_{coll} = \emptyset$, (2) Stage 2 jointly reports $V_{adj}$ and $V_{xtalk}$. This joint reporting is critical: these sets impose opposing spatial pressures—adjacency requires proximity, whereas crosstalk requires separation. Addressing them in isolation causes the model to over-correct in a single direction, leading to oscillatory behavior that hinders convergence. By exposing these \textit{conflicting spatial constraints} simultaneously, the model can navigate the trade-off and identify balanced placements that satisfy all requirements.

\subsection{Routing Agent}
\begin{figure*}[tp]
    \centering
    \includegraphics[width=\linewidth, trim=0in 0in 0in 0in, clip]{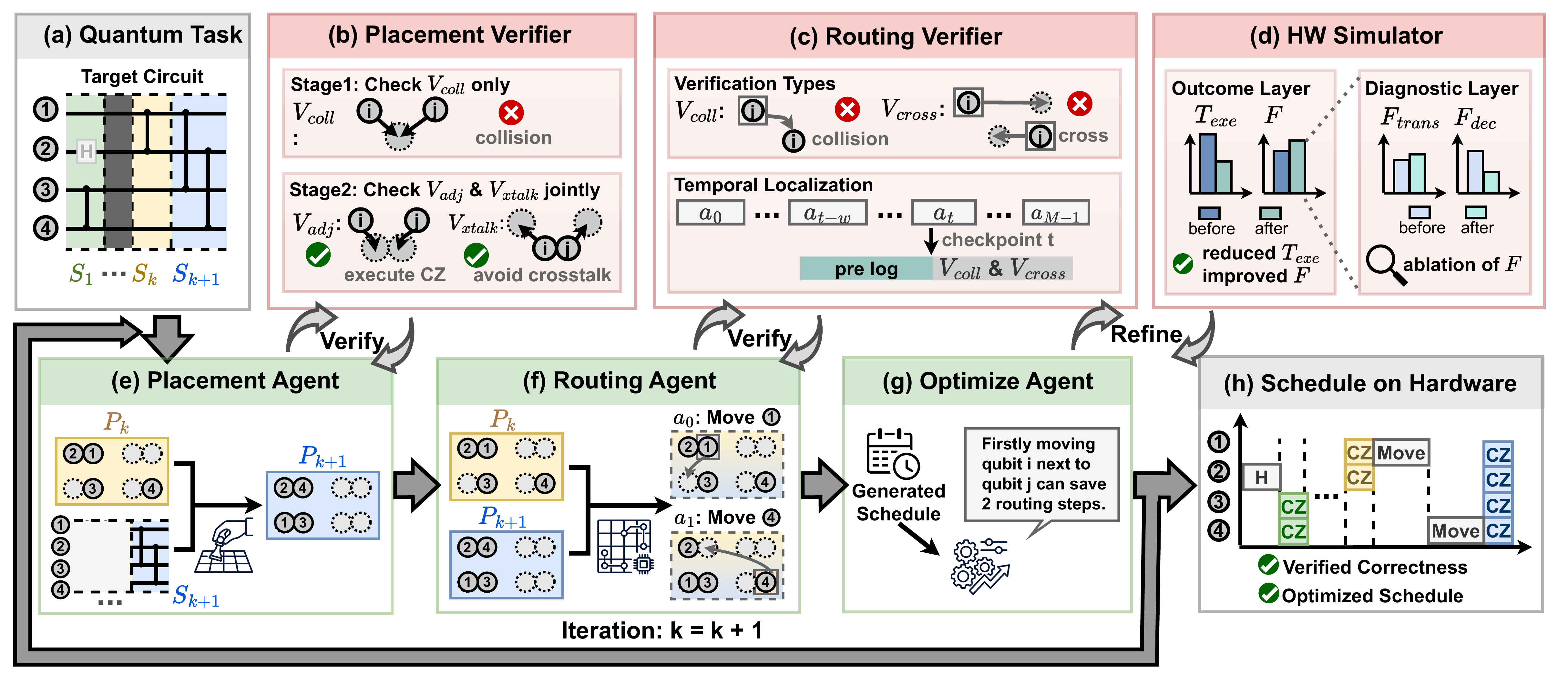}
        \vspace{-6pt}
    \caption{\textbf{Multi-agent framework for quantum compilation.} LLM agents (e–g, green) generate placements, routes, and optimizations; deterministic verifiers (b–d, red) enforce constraints and evaluate performance. Agents iteratively refine outputs via structured feedback until all constraints are satisfied, producing a verified, optimized schedule (h).}
    \label{fig:overview}
\end{figure*}

The Routing Agent (Fig.~\ref{fig:overview}(f)) determines a sequence of $M$ discrete AOD operations $\mathcal{A} = \{a_0, \dots, a_{M-1}\}$ to shuttle atoms for the $(k+1)$-th stage. We denote the mapping after the $t$-th routing step as $P_{k+1}^t$, where $P_{k+1}^0$ is the initial configuration and $P_{k+1}^M$ is the target placement. Each action $a_t$ transitions the system state from $P_{k+1}^t$ to $P_{k+1}^{t+1}$.

\textbf{Spatio-temporal verification and feedback.} To ensure valid trajectories, the verifier evaluates each state transition $P_{k+1}^t \xrightarrow{a_t} P_{k+1}^{t+1}$ against AOD hardware constraints.

\textit{1) Spatio-temporal violation localization.} Routing violations are localized to the offending qubits and the specific step $t$. The verifier produces two structured violation sets of tuples $(q_i, q_j, P_{k+1}^t, P_{k+1}^{t+1}, t)$:
\begin{itemize}
    \item Collision ($V_{coll}$): This violation occurs when multiple qubits are assigned to the same trap position after an operation, i.e., $P_{k+1}^{t+1}(q_i) = P_{k+1}^{t+1}(q_j)$.
    \item Crossing ($V_{cross}$): This violation represents a breach of the rigid AOD ordering, occurring if action $a_t$ causes the relative row or column indices of $q_i$ and $q_j$ in $P_{k+1}^{t+1}$ to invert compared to their order in $P_{k+1}^t$.
\end{itemize}

\textit{2) Temporal localization as causal signal.} Incorporating the step index $t$ within each violation tuple provides a critical causal signal for trajectory refinement. For instance, a collision at step $t=2$ indicates that the initial departure strategy is flawed, whereas a collision at $t=M-1$ suggests that only the final approach to the target $P_{k+1}$ requires adjustment. Without this temporal context, the model lacks the granularity to distinguish between these distinct failure modes and often defaults to redundant global regenerations. By localizing violations to specific transitions $P_{k+1}^t \xrightarrow{a_t} P_{k+1}^{t+1}$, the agent can surgically modify the offending segments while preserving the valid portions of the routing sequence.
The complete violation report, $V = V_{coll} \cup V_{cross}$, is serialized into natural language and appended to the agent's context for iterative refinement. Upon successful verification, the final mapping $P_{k+1}^M$ is established as the initial configuration $P_{k+2}^0$ for the subsequent gate stage $S_{k+2}$.

\subsection{Optimize Agent}

The initial compilation phase ensures architectural correctness but does not guarantee an optimal schedule. The Optimize Agent (Fig.~\ref{fig:overview}(g)) iteratively refines the schedule by sampling candidate modifications and evaluating them against the Hardware Simulator (Fig.~\ref{fig:overview}(d)).

\paragraph{The pitfall of scalar rewards.} 
The simulator evaluates candidates across four dimensions: hardware runtime ($T_{\text{exe}}$), fidelity ($F$), transfer penalty (SLM-AOD error), and decoherence (idle error). A conventional approach would collapse these into a single scalar reward. However, $T_{\text{exe}}$ and $F$ are coupled via competing physical mechanisms. A faster schedule reduces decoherence but often requires aggressive routing that increases transfer overhead. A scalar reward masks which mechanism dominates, leaving the LLM unable to discern whether a fidelity drop stems from excessive transfers or decoherence. These two failure modes require diametrically opposite corrections.

\paragraph{Outcome-diagnostic decomposition.} 
To resolve this ambiguity, the simulator provides a two-layer feedback structure that exposes the underlying causal chain:
\begin{itemize}
    \item Outcome Layer: Reports $T_{\text{exe}}$ and $F$ to characterize global performance trade-offs.
    \item Diagnostic Layer: Decouples $F$ into transfer penalty and decoherence components.
\end{itemize}
This decomposition allows the agent to pinpoint the root cause of performance shifts. For instance, if a faster schedule degrades fidelity, the diagnostic layer can reveal that while decoherence improved, the transfer penalty increased disproportionately. Without this granularity, the LLM might naively attempt to slow down the schedule to improve fidelity. Such a move would further exacerbate decoherence, resulting in the worst possible physical correction.
A candidate modification is accepted if and only if it improves at least one of $T_{\text{exe}}$ or $F$ without degrading the other. If no candidate in a round meets this criterion, the proposal is rejected and the current best schedule is retained. Crucially, diagnostic feedback from rejected candidates is not discarded. It serves as a negative constraint that narrows the search space for subsequent rounds.

\section{Convoy-style Shuttling}\label{sec:case_study}

\begin{figure*}[tp]
    \centering
    \includegraphics[width=\linewidth, trim=0in 0.1in 0in 0in, clip]{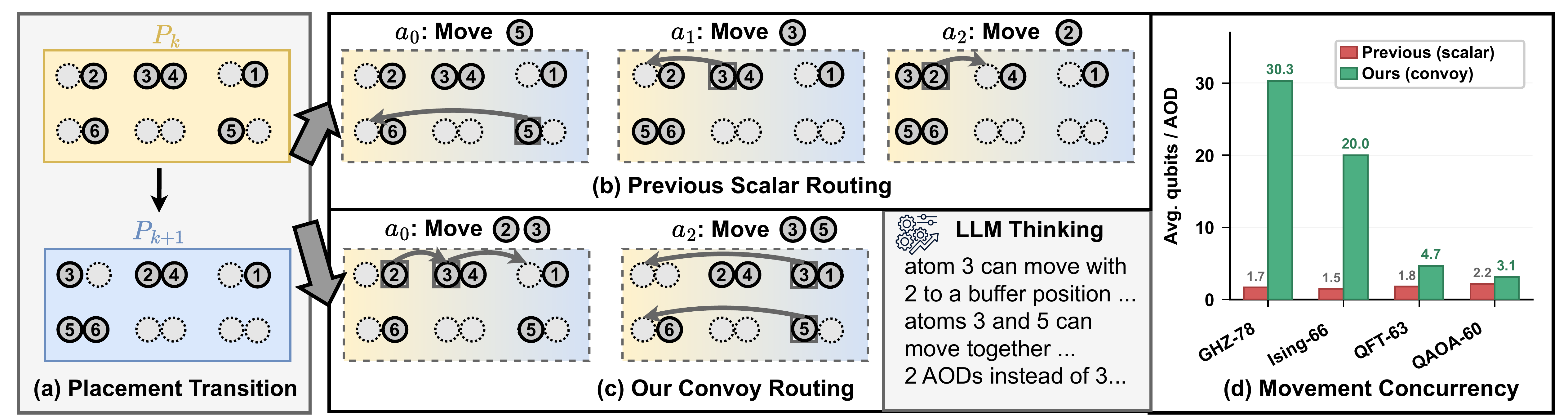}
        \vspace{-6pt}
    \caption{\textbf{Convoy routing overview.} (a) Placement transition between two CZ stages. (b) Scalar routing: qubits are moved individually. (c) Convoy routing: multiple qubits are moved in lockstep, reducing hardware runtime and decoherence. (d) Movement concurrency across benchmarks.}
    \label{fig:convoy}
\end{figure*}

Applying the framework described above, we identify a novel routing pattern that substantially improves hardware schedule efficiency, which we call \emph{convoy-style shuttling}. Unlike previous compilers that typically route qubits independently toward their target traps along shortest paths (Fig. \ref{fig:convoy}(b)), our agent groups qubits with spatially compatible trajectories to displace them in a single, coordinated movement (Fig. \ref{fig:convoy}(c)). By analyzing the LLM's thinking logs, we found that the formation of this pattern is primarily driven by two distinct strategies:



(1) Strategic use of buffer positions: Crucially, the agent often utilizes buffer positions to perform subtle local adjustments. By intentionally deviating from individual shortest paths, the agent aligns qubits whose movements would otherwise be incompatible, allowing them to form a shared convoy. Since AOD rows and columns move in tandem, this deliberate synchronization effectively converts idle hardware cycles into concurrent movements without introducing crossing or collision violations.

(2) Minimizing transfer penalties: The agent discovers that maintaining atoms in continuous motion across consecutive AOD rounds avoids the penalties associated with repeated SLM–AOD transfers. Effectively, the agent converts the physical fidelity model into a scheduling constraint that favors sustained convoy membership over naive round-by-round re-optimization.

Together, these two strategies enable a highly coordinated execution that dramatically increases AOD concurrency (Fig. \ref{fig:convoy}(d)), transforming disjointed routing steps into synchronized, penalty-minimizing convoys.
\toSonia{Specifically, the gap between convoy and scalar routing is most pronounced on circuits with regular gate structures (GHZ, Ising) where compatible displacement vectors arise naturally. While this advantage reduces on irregular topologies (QFT, QAOA) where spatial locality is scarce, the agent maintains a higher concurrency floor than previous path-finding across all benchmarks.}

This discovery is notable in three aspects. (1) The LLM discovers parallelization strategies that facilitate convoy-style shuttling without explicit heuristic guidance. (2) These strategies are non-trivial, requiring joint reasoning over AOD constraints, transfer penalties, and decoherence, a trade-off space that conventional compilers navigate with hand-tuned cost functions. (3) The pattern is directly actionable: the utilization of buffer positions and continuous motion can be extracted as a standalone heuristic for future compilers, illustrating that LLM-driven search can surface optimization opportunities that complement existing domain expertise.
\section{Evaluation}\label{sec:evaluation}

\subsection{Experiment Setup}\label{subsec:setup}

\paragraph{LLM configuration.}
Qubrio does not depend on a specific LLM backbone and uses LLMs only as proposal generators: every candidate placement, route and optimization is accepted only after deterministic verification and simulator evaluation. We evaluate three representative LLM backbones (Qwen-3.5 \cite{qwenteam2026qwen35omnitechnicalreport}, DeepSeek-V3.2 \cite{deepseekai2025deepseekv32pushingfrontieropen} and GPT-5.4 \cite{openai2026gpt54}), spanning closed- and open-source frontier models, to assess backbone sensitivity. Detailed parameters, retry budgets, and run-to-run variance are reported in Appendix \ref{app:raw_data}.

\paragraph{Benchmarks.}
Following the prior work evaluation \cite{ruan2025powermove,bochen2024compilation}, we select a variety of benchmark programs from QASMBench \cite{li2023qasmbench}, including QFT, QAOA, GHZ, and Ising, across multiple circuit sizes (up to the order of $10^4$ gates). They cover the most widely used algorithmic and state-preparation primitives in near-term quantum computing, spanning a wide range of \todo{qubit interaction topologies}.


\paragraph{Baselines.}
We compare against Enola~\cite{bochen2024compilation} and PowerMove~\cite{ruan2025powermove}, which is the SOTA NA compiler. All compilers are evaluated under the
same hardware configurations following prior work \cite{bochen2024compilation, ruan2025powermove}. 
We use PowerMove as the heuristic backbone for our ablation study.



\paragraph{Metrics.}
We report two primary metrics: (1)~\emph{Hardware runtime}~($T_\text{exe}$), \toSonia{the end-to-end execution time of the entire circuit on NA hardware, integrating gate durations, atom movement, and SLM–AOD transfer overheads.}
(2)~\emph{Circuit fidelity}~($F$), the end-to-end probability that a benchmark completes without any logical error, which \todo{jointly} 
account for gate fidelity, transfer penalties and decoherence.  Both metrics are evaluated using the NA \todo{hardware simulator} 
from PowerMove and Enola~\cite{bochen2024compilation, ruan2025powermove} with identical parameter settings as described in Appendix \ref{app:simulator}. \toSonia{Hardware feasibility is enforced as a hard constraint by the deterministic verifiers. Schedules failing verification are reported as `-'.}


\subsection{Main Results}

\paragraph{Comparison to heuristic compilers.}

\begin{figure*}[t]
    \centering
    \includegraphics[width=0.95\linewidth]{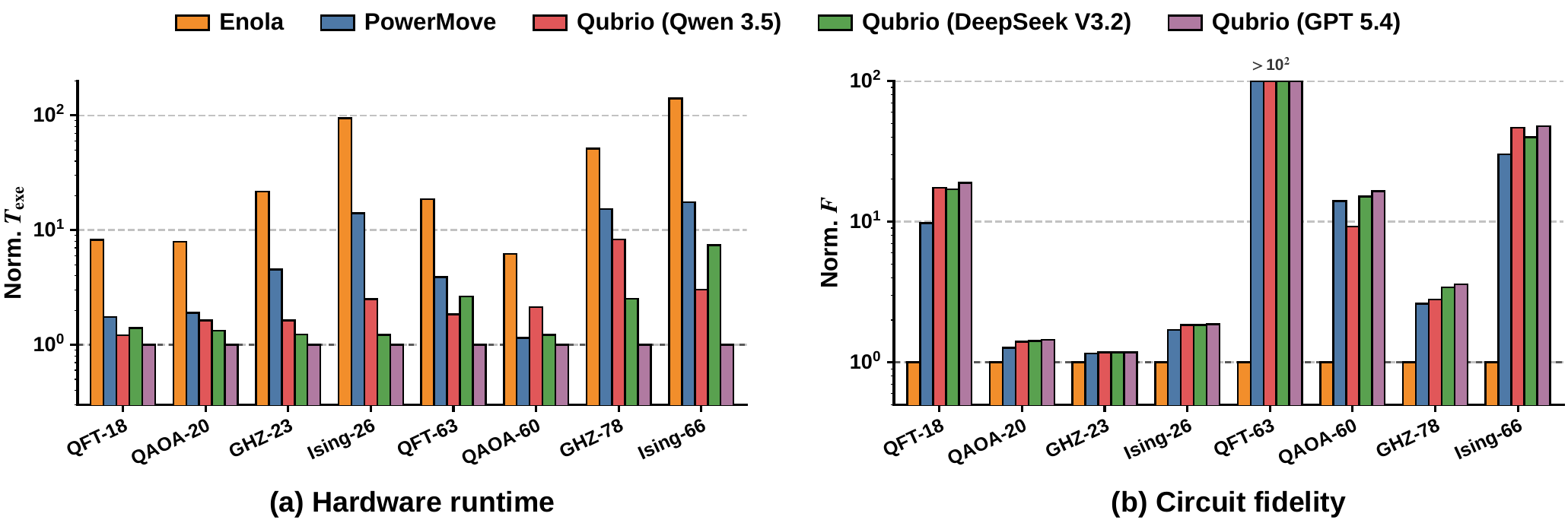}
        \vspace{-5pt}
    \caption{%
        \textbf{End-to-end comparison of Qubrio against Enola and PowerMove.}
 \textbf{(a)} Hardware runtime, normalized to Qubrio (GPT-5.4). \textbf{(b)} Circuit fidelity, normalized to Enola.
    }
    \label{fig:main_results}
\end{figure*}

\emph{(1) Hardware runtime.}
As shown in Fig.~\ref{fig:main_results}(a), Qubrio\,(GPT-5.4) achieves the lowest hardware runtime across all benchmarks. On $\sim$20-qubit circuits, 
Qubrio\,(GPT-5.4) reduces hardware runtime by an
average of \todo{19.1$\times$}/\todo{3.8$\times$} versus Enola/PowerMove. This gap widens at \todo{$\sim$60-qubit circuits}, 
where Qubrio\,(GPT-5.4) reduces hardware runtime by an
average of 
\todo{30.3$\times$}/\todo{5.9$\times$} versus Enola/PowerMove.
The performance advantage stems primarily from convoy-style shuttling, \toSonia{which packs more qubits into each AOD movement round and reduces the total number of rounds (Fig.~\ref{fig:convoy}(d)), directly shortening the movement duration.} As discussed in Section~\ref{sec:case_study}, the variation in performance gap across benchmarks stems from their circuit structural differences. \toSonia{On regular circuits (GHZ, Ising) characterized by local gate interactions, Qubrio (GPT-5.4) achieves an average speedup of $11.4\times$ over PowerMove. On irregular circuits (QFT, QAOA) with non-local and dense interaction patterns, the average speedup reduces to $2.0\times$.}
Qubrio\,(DeepSeek-V3.2) and Qubrio\,(Qwen-3.5) also
outperform Enola/PowerMove by an average of
\todo{12.7$\times$}/\todo{2.5$\times$} and
\todo{10.4$\times$}/\todo{2.1$\times$}.
The better performance of both LLM variants suggests that these gains are inherent to the agentic architecture rather than being contingent on the reasoning capabilities of a specific LLM model. 

\emph{(2) Circuit fidelity.}
As shown in Fig.~\ref{fig:main_results}(b), Qubrio (GPT-5.4) achieves higher fidelity across all benchmarks. The most significant gain is on QFT-63, where massive global routing penalties cause Enola and PowerMove to collapse to $10^{-69}$ and $10^{-16}$. Qubrio recovers this to $10^{-8}$ by effectively managing dense connectivity through its efficient convoy-style shuttling. Excluding this outlier, Qubrio still improves fidelity by an average of $5.6\times$/$1.30\times$ over Enola/PowerMove (reaching a $9.9\times$ improvement over PowerMove when included). These gains naturally stem from our reduced hardware runtimes and transfer penalties, which minimize the primary sources of hardware errors. Furthermore, Qubrio\,(DeepSeek-V3.2) and Qubrio\,(Qwen-3.5) also demonstrate strong performance, improving fidelity by $5.2\times$/$1.22\times$ and $4.9\times$/$1.13\times$ (excluding QFT-63). Across all underlying LLMs, these fidelity improvements match the conclusions drawn from our hardware runtime analysis.


\emph{(3) Cost analysis.}
The financial overhead of LLM-driven compilation is negligible relative to \HZ{QPU} execution expenses and is rapidly amortized by the resulting QPU runtime savings. Specifically, compiling via DeepSeek-V3.2 incurs per-stage costs from \$0.29 ($\sim$20 qubits) to \$0.59 ($\sim$60 qubits). 
At current NA QPU pricing \cite{teegarden2026monthslifecloudquantum, tqi2025pricing}, a standard execution costs roughly \$10. Since QPU usage scales with hardware runtime, the runtime reduction delivered by Qubrio saves a substantial fraction of this cost per execution, so the one-time compilation expense is offset within a few standard executions, after which all subsequent runs yield net savings.
By compiling batches of stages in parallel (Appendix \ref{app:parallel}), end-to-end compilation takes a scale-invariant ${\sim}$5 minutes. This latency is a one-time cost: the generated routing schedules are reusable, requiring no recompilation for different inputs or local parameter updates \cite{zhao2026system, krol2023efficient, dalvi2024one}.



\begin{figure*}[tp]
    \centering
    \includegraphics[width=0.95\linewidth]{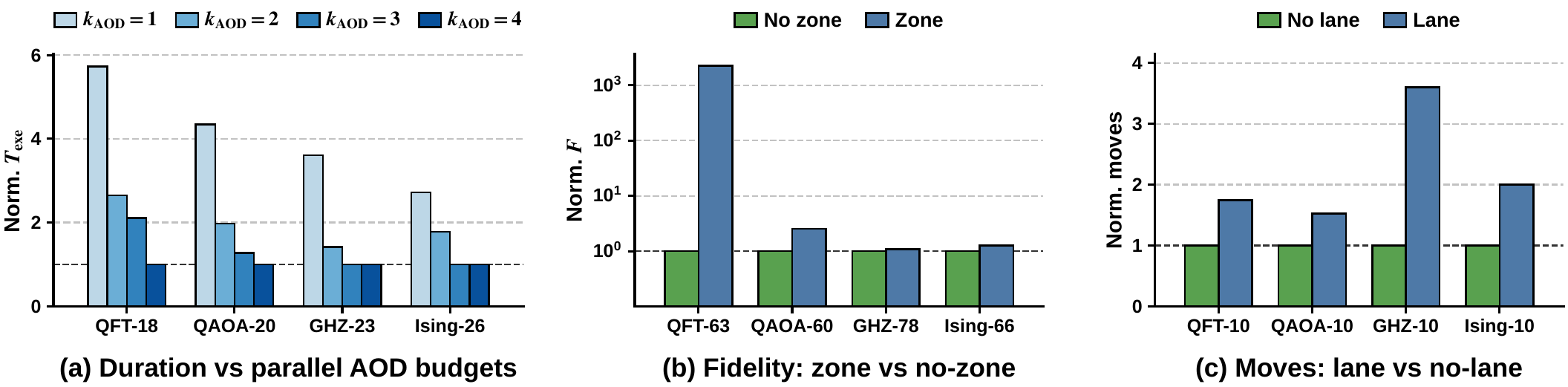}
        \vspace{-5pt}
    \caption{%
\textbf{Adaptability to three hardware capabilities.} \textbf{(a)} Hardware runtime normalized to the $k_{\mathrm{AOD}} = 4$ case across different parallel AOD budgets ($k_{\mathrm{AOD}} \in \{1,2,3,4\}$). \textbf{(b)} Fidelity comparison between zoned and non-zoned scenarios, normalized to the non-zoned case.\textbf{(c)} \todo{ Number of atom movements on a lane architecture, normalized to the no-lane case.}
    }
    \label{fig:adaptability}
\end{figure*}

\paragraph{Adaptability to three hardware capabilities.}
Fig.~\ref{fig:adaptability} demonstrates Qubrio's adaptability to three hardware capabilities, using 
\todo{DeepSeek-V3.2} 
as an example backbone model.

\emph{(1) Adapt to multi-AOD scenario.}
Supporting multiple parallel AODs in a heuristic compiler requires extending the router from single-AOD pathfinding to multi-AOD assignment and scheduling, along with rederiving the associated cost model. 
As shown in Fig.~\ref{fig:adaptability}(a), a single additional AOD already yields substantial duration reductions across all benchmarks, with further gains as $k_\text{AOD}$ increases. 
This demonstrates that the LLM agent can effectively leverage \todo{modest} hardware additions to substantially improve execution time, without any change to the underlying compilation logic.

\emph{(2) Adapt to zoned architecture.}
\todo{Zoned Architectures} partition the trap array into spatial regions with distinct operational roles \cite{bluvstein2024logical}. Previous support requires modifying placement to respect zone boundaries and constraining the router to avoid illegal cross-zone transfers. 
Fig.~\ref{fig:adaptability}(b) shows that the fidelity under zoned architectures consistently outperforms that of the non-zoned case across circuits, with a notable $2263\times$ improvement on QFT-63.
This gain arises because zone partitioning eliminates atoms' exposure to excitation error during two-qubit gate execution,
and the agent exploits this structure to improve fidelity.

\emph{(3) Adapt to lane architecture.}
We further evaluate Qubrio on the recently introduced lane architecture by QuEra \cite{luo2026building}, which reduces AOD transport calibration overhead by restricting atom movements to a predefined set of calibrated transport lanes. While this design improves hardware control efficiency, the reduced transport flexibility introduces additional routing challenges, as atoms may need multiple hops to reach their destinations.
As shown in Fig. \ref{fig:adaptability}(c), Qubrio successfully adapts to this more constrained architecture while maintaining reasonable execution performance. Although the number of atom movements increases compared with unrestricted transport, the overhead remains within a manageable range, demonstrating Qubrio's adaptability to hardware architectures with different transport capabilities.

Across these hardware scenarios, adapting Qubrio follows a lightweight and consistent procedure. Specifically, only three components require updates: (1) the hardware description in the LLM prompt, (2) the architecture-specific legality constraints enforced by the verifier, and (3) the hardware parameters and supported operations modeled by the simulator. These changes can be directly derived from the corresponding hardware specifications without modifying the core compilation algorithms, enabling Qubrio to accommodate diverse hardware capabilities with minimal manual effort.

\subsection{Ablation Study}\label{sec:ablation}

\begin{figure*}[tp]
    \centering
    \includegraphics[width=0.95\linewidth]{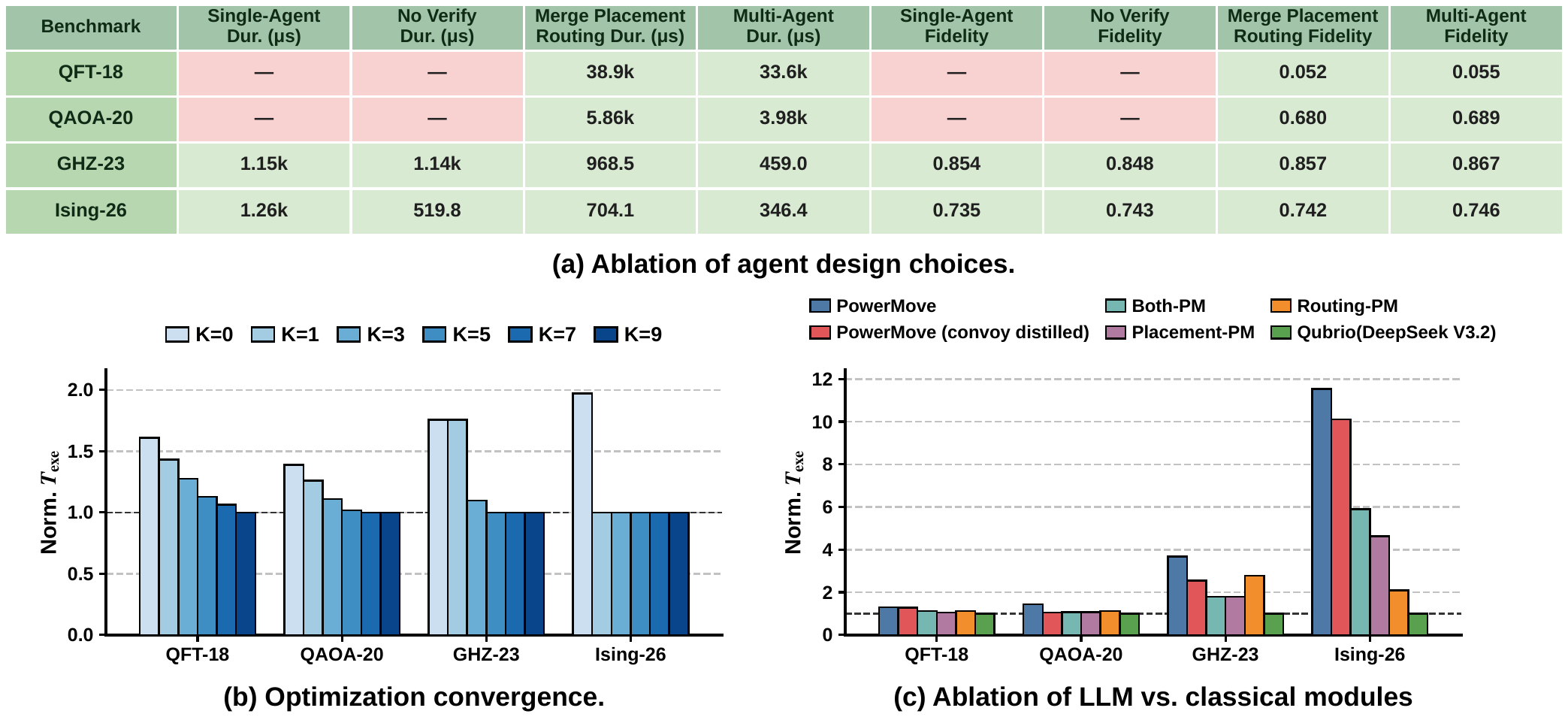}
    \vspace{-5pt}
    \caption{%
        \textbf{Ablation studies isolating architectural contributions. }
    }
    \label{fig:ablation}
\end{figure*}

\paragraph{Ablation of agent design choices.} 
Table~\ref{fig:ablation}(a) isolates the contribution of each architectural decision by comparing four pipeline configurations. \HZ{(1)} Single-Agent, which removes both stage-wise decomposition and the placement-routing separation, fails to produce valid schedules on QFT-18 and QAOA-20, indicating that a single LLM pass overwhelms the context window and conflates distinct reasoning modes. \HZ{(2)} No-Verifier also fails on these two circuits, showing that stage-wise decomposition alone is insufficient without closed-loop constraint correction. On GHZ-23 and Ising-26 where both do converge, their schedules are \todo{2.4$\times$} longer than the full pipeline. \HZ{(3)} Merge Placement Routing, which retains the verifier but merges placement and routing into a single agent, succeeds on all four circuits, demonstrating that the verifier is the critical ingredient for reliability. (4) Multi-Agent, which decouples placement and routing into separate agents, further reduces duration by an average of \todo{1.6$\times$} across the board while maintaining comparable fidelity. These results confirm that agent decomposition and the verification mechanism are complementary: the verifier ensures correctness, while decomposition allows agents to reason effectively within focused scopes. 

\paragraph{Sensitivity to optimization rounds.}
Fig.~\ref{fig:ablation}(b) shows schedule duration as the number of optimizer turns $K$ 
is swept over \todo{$\{0,1,3,5,7,9\}$}, min-normalized to $K$=\todo{9}. Without optimization ($K$=0), schedules are on average \todo{1.7$\times$} longer. The \todo{fifth} round ($K$=\todo{5}) captures the bulk of the improvement. This convergence suggests that the disaggregated feedback structure enables the LLM to identify high-impact modifications early rather than requiring many rounds of trial and error.

\paragraph{Ablation of LLM vs.\ heuristic modules.}
Each LLM module contributes to the overall pipeline performance. As shown in Fig.~\ref{fig:ablation}(c), progressively replacing LLM agents with heuristic PowerMove counterparts across five configurations, while retaining the verifier, increases hardware runtime. The LLM advantage is moderate on irregular circuits (QFT-18, QAOA-20) and grows on regular structures, reaching an $11.5\times$ improvement on Ising-26.
Furthermore, the routing stage benefits more from LLM reasoning than the placement stage. Comparing the single-replacement variants shows that using a heuristic for routing degrades performance more than using a heuristic for placement on three of the four benchmarks. This is likely because sequential trajectory planning under coupled AOD constraints involves a large combinatorial search space, where the LLM's ability to evaluate multi-step trade-offs provides an advantage over heuristic search.
Finally, the discovered convoy-style shuttling strategies demonstrate practical transferability. Integrating these strategies into the PowerMove baseline recovers 38\% of the total runtime reduction. The remaining performance gap aligns with our expectations: while hardcoded rules effectively establish coarse-grained parallel routing patterns, they lack the flexibility to  handle granular, circuit-specific decisions. 

\begin{figure*}[tp]
    \centering
    \includegraphics[width=0.95\linewidth]{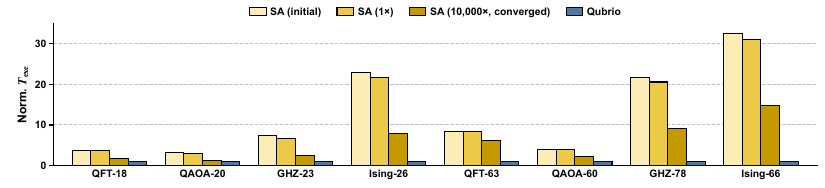}
    \vspace{-5pt}
    \caption{%
        \textbf{Comparison with Simulated Annealing Across Different Evaluation Budgets. }
    }
    \label{fig:sa}
\end{figure*}

\paragraph{Contribution of LLM Reasoning.}
To evaluate the contribution of LLM reasoning, we replace both LLM agents in Qubrio with simulated annealing (SA), a standard combinatorial optimizer \cite{aarts1989simulated}, while retaining the same verifiers and hardware simulator. Fig. \ref{fig:sa} compares SA under two simulator-evaluation budgets: SA (1×), which matches Qubrio's evaluation budget, and SA (10,000×), at which SA converges across all benchmarks. We sweep the initial acceptance probability following \cite{aarts1989simulated} and report the best configuration for each benchmark.

At the matched evaluation budget, SA achieves only limited improvements over its initial schedule. Even with 10,000× more evaluations, Qubrio consistently achieves lower hardware execution time and higher circuit fidelity across all benchmarks. These results indicate that Qubrio's improvements cannot be attributed solely to additional search or simulator evaluations. Instead, they highlight the value of LLM-guided reasoning in navigating the coupled placement and routing constraints, where conventional search-based optimization can become trapped in poor local optima \cite{yang2026unifying,zou2024lightsabre}.

\section{Related Work}

\textbf{Quantum Compilation.}
Existing frameworks address qubit routing across diverse platforms, including superconducting devices~\cite{qiskit, burgholzer2025mqt, cowtan2019qubit, li2019tackling, zulehner2018efficient, tan2020optimal, pattanayak2021quantum, sivarajah2021t}, photonics~\cite{zhang2023oneq, zhang2024oneperc, zhang2025oneadapt}, ion traps~\cite{murali2020architecting, ruan2025trapsimd,zhu2025s,bach2025efficient}, and multi-chip systems~\cite{bravyi2022future, wang2016chip, zhang2024mech, wu2022autocomm}. Among these, NA architectures present a uniquely challenging compilation target~\cite{wang2024atomique, wang2024q, tan2022qubit, lin2024reuseawarecompilationzonedquantum, tan2024compiling, bochen2024compilation, ruan2025powermove}. NA qubits are shuttled via AODs restricted by non-crossing constraints~\cite{bluvstein2022quantum, bluvstein2024logical}, creating a globally coupled trajectory planning problem governed by velocity limits and transfer-decoherence trade-offs. Current NA compilers~\cite{bochen2024compilation, ruan2025powermove} navigate this space using hand-engineered heuristics, meaning adaptations for hardware features like parallel AODs require redesigning routing logic and cost models~\cite{ruan2025powermove}. 


\textbf{Machine Learning for Quantum Compilation.}
Recent ML advances in quantum compilation, such as RL for SWAP routing~\cite{pozzi2022using, tang2024alpharouter, ngo2025fiddle} and GNNs for qubit allocation~\cite{lecompte2023graph, sang2026learningoptimizedqubitmappingreuse}, are tailored for superconducting platforms where routing is local on fixed graphs and single-step errors are not strictly fatal.
In contrast, NA hardware introduces a non-local, highly constrained routing paradigm. A single AOD operation shifts multiple atoms simultaneously, enforcing rigid no-crossing constraints across all trajectories. This global mechanical linkage invalidates standard ML: RL suffers from prohibitively sparse rewards since a single collision breaks the entire trajectory, while GNNs' local message-passing fails to capture non-local AOD row and column dependencies. Therefore, NA routing demands a paradigm capable of global reasoning and complex constraint satisfaction rather than step-by-step local optimization.
While LLMs possess the requisite advanced reasoning for such tasks, existing quantum LLM research sidesteps hardware constraints to focus on algorithmic circuit generation~\cite{jern2025agent, liang2023unleashing, sinha2025circuit} instead of compilation onto hardware. No existing ML-based method has demonstrated end-to-end compilation for NA hardware, and state-of-the-art NA compilers rely mostly on hand-engineered heuristics~\cite{bochen2024compilation, 
ruan2025powermove}.

\section{Conclusion}\label{sec:conclusion}
We presented Qubrio, an agentic framework for NA quantum compilation. By deploying specialized agents grounded by deterministic verifiers and disentangled simulator feedback, Qubrio resolves core LLM bottlenecks in quantum compilation. Consequently, Qubrio delivers average improvements of \todo{4.7$\times$} and \todo{1.3$\times$} in hardware runtime and fidelity over the SOTA compiler. The framework also seamlessly adapts to parallel AODs, zoned architectures and the laned architecture without algorithmic changes and discovers novel strategies that facilitate convoy-style shuttling, substantially enhancing hardware concurrency without sacrificing fidelity. Current evaluations focus on NA hardware. Future work could be empirically validating this framework on other quantum platforms.


\section{Acknowledge}\label{sec:ack}
This work was supported in part by the U.S. National Science Foundation (NSF) under Grants No. 2552449, 2534655, 2547483, 2427109, and 2422169, and the U.S. Department of Energy (DOE) under Award No. DE-SC0026692. We also thank QuEra Computing for their valuable guidance on hardware-related aspects of this work.
\newpage
\bibliographystyle{unsrtnat}
\bibliography{reference} 

\newpage
\appendix

\section{Agent Prompts and Output Schemas}
\label{app:prompts}

This appendix summarizes the prompt templates used by Qubrio's LLM agents and connects the abstract agent descriptions in the main text to their implementation. The implementation contains four prompt templates: initial placement generation, stage placement generation, AOD transition routing, and schedule optimization. These templates correspond to the three agents described in the main text: the Placement Agent, Routing Agent, and Optimize Agent. Across all templates, the LLM is used as a proposal generator; generated placements, routes, and optimized schedules are accepted only after deterministic validation.

\vspace{0.3em}
\noindent \textbf{Placement Agent.}
Qubrio uses two placement prompts. The first generates the initial qubit-to-trap mapping before any routing is required, while the second generates the target layout for each subsequent CZ stage. Both prompts output a complete logical-to-physical mapping and are checked by the Placement Verifier before routing begins.

\emph{Initial placement.}
The initial placement prompt specifies the hardware grid width, height, and number of qubits. Trap coordinates are integer lattice points, each physical trap can host at most one atom, and each gate site consists of a left trap and a right trap defined by
\[
\mathrm{group\_id}(x,y)=\mathrm{HEIGHT}\cdot y+\lfloor x/2 \rfloor .
\]
The prompt provides the first CZ stage and asks the model to assign every qubit to exactly one valid trap coordinate. The main optimization instruction is dense clustering: atoms should be packed into a compact rectangular or square block, rather than spread randomly across the entire grid, to reduce future movement distance.

The required output schema is:
\begin{verbatim}
{
  "initial_layout": {
    0: (x, y),
    1: (x, y),
    ...
  }
}
\end{verbatim}

\emph{Stage placement.}
The stage placement prompt receives the current layout, the next CZ stage, and a lookahead list of future CZ stages. For the immediate stage, each CZ pair must occupy the same gate site, with one atom in the left trap and one atom in the right trap. This same-gate-site condition is the implementation-level realization of the interaction-distance constraint used in the main text. The prompt also allows optional prefetching: qubits that interact in future stages may be moved closer in the current stage, as long as the immediate CZ constraints remain satisfied.

The prompt enforces complete assignment, trap capacity, and valid CZ-pair placement constraints. 

The required output schema is:
\begin{verbatim}
{
  "target_layout": {
    0: (x, y),
    1: (x, y),
    ...
  }
}
\end{verbatim}

\vspace{0.3em}
\noindent \textbf{Routing Agent.}
The Routing Agent receives the current layout and the target layout produced by the Placement Agent. It outputs a sequence of AOD move rounds that physically route atoms to their target traps. Each move round is a set of simultaneous atom movements.

The routing prompt specifies two hard physical constraints. First, the capacity constraint requires that, at the end of each round, no physical trap contains more than one atom, accounting for atoms that leave and arrive in the same round. Second, the no-crossing symbolic rule requires that, within a single AOD round, the strict relative ordering of moving atoms along both the $x$ and $y$ axes must not flip.

The required output schema is:
\begin{verbatim}
{
  "aod_move_groups": [
    [ (q, (x_src, y_src), (x_dst, y_dst)), ... ],
    [ (q, (x_src, y_src), (x_dst, y_dst)), ... ],
    ...
  ]
}
\end{verbatim}

\vspace{0.3em}
\noindent \textbf{Optimize Agent.}
The Optimize Agent refines an already valid AOD schedule. It receives the current layout, the target layout, vacant traps, hardware dimensions, and an unoptimized list of AOD move groups. The model is instructed not to change the final target layout, but only to improve the route used to reach it.

The prompt enforces the same hard physical constraints as the Routing Agent: no crossing along either axis, valid trap occupancy at arrival, and unique destinations within each round. 

For iterative optimization, the prompt receives two-layer feedback from the Hardware Simulator. The outcome layer reports hardware runtime $T_{\mathrm{exe}}$ and circuit fidelity $F$. The diagnostic layer decomposes fidelity into transfer fidelity and coherence fidelity, allowing the model to distinguish whether a candidate schedule is limited by excessive SLM--AOD transfers or excessive idle decoherence. A candidate is accepted only if it improves at least one of $T_{\mathrm{exe}}$ or $F$ without degrading the other.

The required output is a JSON array of optimized AOD move groups:
\begin{verbatim}
[
  [ [qubit_id, [start_x, start_y], [end_x, end_y]], ... ],
  ...
]
\end{verbatim}

\section{Deterministic Verifiers}
\label{app:verifier}

Qubrio uses LLMs as proposal generators and relies on deterministic verifiers to enforce hardware validity. This appendix describes the implementation-level checks corresponding to the Placement Verifier and Routing Verifier in the main text. Invalid outputs are rejected and converted into feedback for subsequent refinement attempts.

\vspace{0.3em}
\noindent \textbf{Placement Verifier.}
The Placement Verifier is implemented by \texttt{diagnose\_layout} in \texttt{verify.py}. Given a hardware instance, a candidate layout, and the current CZ stage, it diagnoses placement violations as
\[
\{V_{\mathrm{coll}}, V_{\mathrm{adj}}, V_{\mathrm{xtalk}}, \mathrm{fatal}\}.
\]
The hardware layout follows the split-trap model: for an $n$-qubit program, the verifier uses
\[
\mathrm{WIDTH}=2\lceil\sqrt{n}\rceil,\qquad
\mathrm{HEIGHT}=\lceil\sqrt{n}\rceil .
\]
A coordinate $(x,y)$ is valid only if
\[
0 \le x < \mathrm{WIDTH}, \qquad 0 \le y < \mathrm{HEIGHT}.
\]
Each gate site is defined by
\[
\mathrm{group\_id}(x,y)=\mathrm{HEIGHT}\cdot y+\lfloor x/2\rfloor,
\]
so the two traps with x coordinates $2k$ and $2k+1$ at the same row belong to the same gate site.

The Placement Verifier first checks fatal layout errors, including missing qubits, unknown qubits, and out-of-bound coordinates. It then checks exact trap collisions, where two logical qubits are assigned to the same physical trap. In addition, if more than two atoms occupy the same gate site, all such pairs are reported as collision-class violations. For the current CZ stage, $V_{\mathrm{adj}}$ contains CZ pairs that are not assigned to the same gate site, while $V_{\mathrm{xtalk}}$ contains non-CZ pairs that occupy the same gate site. Each categorized violation is localized as
\[
(q_i, q_j, P(q_i), P(q_j)).
\]
Thus, the interaction-distance and crosstalk constraints in the main text are instantiated by same-gate-site checks in the implementation.

The placement feedback loop is implemented in \texttt{\_format\_placement\_feedback} in \texttt{incremental\_compiler.py}. It follows the two-stage policy described in the main text. If any collision exists, the feedback exposes only $V_{\mathrm{coll}}$, so the model first establishes a valid trap assignment. Once collisions are resolved, the feedback jointly reports $V_{\mathrm{adj}}$ and $V_{\mathrm{xtalk}}$, since these two constraints impose opposing spatial pressures. A placement is accepted only when all violation sets and fatal errors are empty.

\vspace{0.3em}
\noindent \textbf{Routing Verifier.}
The Routing Verifier checks AOD movement rounds. Each AOD round is represented as a list of moves
\[
(q, (x_{\mathrm{src}}, y_{\mathrm{src}}), (x_{\mathrm{dst}}, y_{\mathrm{dst}})).
\]
The online checker is implemented by \texttt{check\_aod\_round} in \texttt{verify.py}. For each round, it enforces: (i) the round is non-empty; (ii) no qubit moves more than once in the same round; (iii) every moved qubit exists in the current layout; (iv) each source coordinate matches the current layout; (v) each destination coordinate is within bounds; and (vi) after the round, each physical trap contains at most one atom. The occupancy check accounts for atoms that leave and arrive in the same round, so vacate-then-occupy moves are permitted when they do not violate final occupancy.

The Routing Verifier also enforces the AOD no-crossing constraint using symbolic ordering. For every pair of moving atoms in the same round, it compares their relative order along the $x$ and $y$ axes before and after the move. A crossing violation is reported if a strict order flips, i.e., $<$ becomes $>$ or $>$ becomes $<$. This check is applied pairwise among all moving atoms in the round.

The implementation also provides a structured diagnostic routine, \texttt{diagnose\_routing} in \texttt{verify.py}, which localizes routing failures as
\[
\{V_{\mathrm{coll}}, V_{\mathrm{cross}}, \mathrm{fatal}\}.
\]
Here, $V_{\mathrm{coll}}$ contains trap-capacity violations after a movement step, and $V_{\mathrm{cross}}$ contains AOD ordering violations. Each categorized routing violation is localized as
\[
(q_i, q_j, P^t, P^{t+1}, t),
\]
where $t$ is the offending movement step. Fatal errors include empty rounds, duplicate movers, unknown qubits, source mismatches, and out-of-bound destinations.

In the compilation loop, routing outputs are parsed and checked in \texttt{generate\_transition} in \texttt{incremental\_compiler.py}. The loop uses \texttt{check\_aod\_round} for online validation and target-layout consistency, while \texttt{diagnose\_routing} provides the structured step-level diagnosis corresponding to the violation sets described in the main text. After all AOD rounds pass validation, the compiler checks that every qubit reaches the target layout produced by the Placement Agent. If any round fails verification or the final layout does not match the target layout, the route is rejected and the error is returned to the LLM as refinement feedback.

\vspace{0.3em}
\noindent \textbf{Optimize Agent validation and acceptance.}
The Optimize Agent refines an already valid AOD schedule. Candidate optimized schedules are parsed in \texttt{optimize\_aod\_moves} in \texttt{incremental\_compiler.py}. Each candidate is replayed from the current layout, and every optimized AOD round is checked by the same deterministic Routing Verifier, \texttt{check\_aod\_round}. The compiler then verifies that the replayed final layout exactly matches the target layout. Therefore, optimization is not allowed to change the target placement; it may only change the route used to reach that placement.

After physical validation, the candidate is evaluated by the Hardware Simulator. The acceptance rule is implemented by \texttt{\_accept\_candidate}: a candidate is accepted only if it improves at least one of hardware runtime or circuit fidelity without degrading the other. Invalid candidates are rejected before simulation. Valid but non-improving candidates are rejected by the acceptance rule. When full feedback mode is enabled, rejected candidates are included in the subsequent optimization context together with outcome-level metrics and diagnostic metrics.

\vspace{0.3em}
\noindent \textbf{Parsing and retry handling.}
LLM outputs are parsed before verification. Placement and routing outputs are extracted as Python dictionaries using \texttt{extract\_python\_dict} in \texttt{incremental\_compiler.py}, then parsed with \texttt{ast.literal\_eval}. Initial placement must contain the \texttt{initial\_layout}, stage placement must contain \texttt{target\_layout}, and routing must contain \texttt{aod\_move\_groups}. The Optimize Agent output is parsed as a JSON array of AOD move groups.

The compiler uses bounded retry loops. Initial placement, stage placement, and routing use an inner retry limit of 10 by default. Optimization uses 5 attempts by default. The full transition loop also has an outer retry limit of 5. When parsing or verification fails, the compiler appends the error message and, when available, a history of failed attempts to the next LLM call. If the retry limit is exhausted, the corresponding compilation stage fails rather than accepting an invalid schedule.

\vspace{0.3em}
\noindent \textbf{Full-schedule sanity check.}
The function \texttt{verify\_output} in \texttt{verify.py} provides a full-schedule sanity check. It validates the initial layout, parses each schedule item, replays all AOD move groups with \texttt{check\_aod\_round}, and audits each CZ stage using \texttt{audit\_groups\_pre\_cz}. The CZ audit checks that each executed CZ stage matches the input stage and that any two atoms sharing a gate site correspond to an intended CZ pair. The verifier also checks that the schedule contains the expected number of CZ stages and that the final layout remains within bounds.
\section{Broader Impacts}
Qubrio has positive societal impacts by reducing the manual effort required to adapt quantum compilers to evolving hardware, potentially accelerating research in quantum computing and making compiler development more accessible. Potential negative impacts include the possibility that more automated and efficient compilation could accelerate dual-use quantum-computing applications in the future, as well as privacy and reproducibility concerns when using commercial LLM APIs.

\section{LLM Configuration and Execution Details}
\label{app:llm_config}

Qubrio employs a model-agnostic architecture with a unified API to support different LLM backbones. In all experiments, the LLM functions solely as a proposal generator; all outputs require deterministic verification and simulator evaluation prior to acceptance.

To assess the impact of the model backbone, we evaluate Qwen-3.5 \cite{qwenteam2026qwen35omnitechnicalreport}, DeepSeek-V3.2 \cite{deepseekai2025deepseekv32pushingfrontieropen} and GPT-5.4 \cite{openai2026gpt54} using their API snapshots accessed in March 2026. By default, the models use temperature 0.7 and top-p 1.0. Verifier-guided correction is capped at 10 rounds for placement and routing, and optimization is capped at 5 rounds.

\section{Compute resources}\label{app:compute}
\label{app:compute_resource}
All verifier and hardware-simulator experiments were run on a CPU server with an Intel Xeon CPU with 32 cores and 256\,GB of RAM. LLM backbones were accessed through external APIs, so no local GPU was required for the reported API-based experiments. Compilation time and per-stage API cost are reported in Section~\ref{sec:evaluation}.
\section{Raw Data for Main Results}\label{app:raw_data}

\begin{table*}[h!]
\centering
\caption{Per-benchmark hardware runtime, circuit fidelity, and LLM retry
counts. Error bars indicate mean ± one standard deviation, computed from N=6 independent runs for each benchmark.}
\label{tab:main_full}
\setlength{\tabcolsep}{4pt}
\renewcommand{\arraystretch}{1.18}
\resizebox{\linewidth}{!}{%
\begin{tabular}{l l rl rl c c}
\toprule
Benchmark & Model
 & \multicolumn{2}{c}{$T_{\mathrm{exe}}$ ($\mu$s)}
 & \multicolumn{2}{c}{$F$}
 & \multicolumn{2}{c}{Average retries} \\
 & & mean & $\pm$std & mean & $\pm$std
 & Placement & Routing \\
\midrule
\multirow{5}{*}{QFT-18}
 & Enola              & $197{,}129.8$ & 0       & $3.26{\times}10^{-3}$ & 0       & N/A & N/A \\
 & PowerMove          & $ 41{,}769.7$ & 0       & $3.18{\times}10^{-2}$ & 0       & N/A & N/A \\
 & Qubrio (Qwen-3.5)  & $ 29{,}007.5$ & $3{,}553$ & $5.66{\times}10^{-2}$ & $7.7{\times}10^{-4}$ & 4.00 & 3.17 \\
 & Qubrio (DS-V3.2)   & $ 33{,}569.1$ & $2{,}954$ & $5.51{\times}10^{-2}$ & $5.5{\times}10^{-4}$ & 1.17 & 3.00 \\
 & Qubrio (GPT-5.4)   & $ 23{,}895.6$ & $1{,}903$ & $6.16{\times}10^{-2}$ & $4.6{\times}10^{-4}$ & 1.00 & 3.00 \\
\midrule
\multirow{5}{*}{QAOA-20}
 & Enola              & $ 23{,}641.9$ & 0       & $0.4864$ & 0       & N/A & N/A \\
 & PowerMove          & $  5{,}690.3$ & 0       & $0.6186$ & 0       & N/A & N/A \\
 & Qubrio (Qwen-3.5)  & $  4{,}887.3$ & $809$   & $0.6844$ & $0.021$ & 2.66 & 3.00 \\
 & Qubrio (DS-V3.2)   & $  3{,}975.3$ & $576$   & $0.6889$ & $0.019$ & 1.33 & 4.00 \\
 & Qubrio (GPT-5.4)   & $  2{,}989.5$ & $173$   & $0.7041$ & $0.018$ & 1.00 & 3.00 \\
\midrule
\multirow{5}{*}{GHZ-23}
 & Enola              & $  8{,}037.5$ & 0       & $0.7377$ & 0       & N/A & N/A \\
 & PowerMove          & $  1{,}684.0$ & 0       & $0.8509$ & 0       & N/A & N/A \\
 & Qubrio (Qwen-3.5)  & $    607.0$   & $59$    & $0.8699$ & $0.003$ & 1.00 & 3.00 \\
 & Qubrio (DS-V3.2)   & $    459.0$   & $39$    & $0.8670$ & $0.003$ & 1.00 & 3.00 \\
 & Qubrio (GPT-5.4)   & $    371.8$   & $62$    & $0.8687$ & $0.001$ & 1.00 & 3.00 \\
\midrule
\multirow{5}{*}{Ising-26}
 & Enola              & $ 27{,}019.2$ & 0       & $0.4035$ & 0       & N/A & N/A \\
 & PowerMove          & $  3{,}992.7$ & 0       & $0.6868$ & 0       & N/A & N/A \\
 & Qubrio (Qwen-3.5)  & $    715.1$   & $127$   & $0.7432$ & $0.004$ & 1.33 & 3.00 \\
 & Qubrio (DS-V3.2)   & $    346.4$   & $61$    & $0.7460$ & $0.003$ & 1.17 & 3.00 \\
 & Qubrio (GPT-5.4)   & $    284.7$   & $145$   & $0.7548$ & $0.005$ & 1.00 & 2.00 \\
\midrule
\multirow{5}{*}{QFT-63}
 & Enola              & $1{,}329{,}226.1$ & 0   & $3.95{\times}10^{-69}$ & 0    & N/A 
 & N/A \\
 & PowerMove          & $   277{,}720.0$  & 0   & $8.02{\times}10^{-16}$ & 0    & N/A & N/A \\
 & Qubrio (Qwen-3.5)  & $   131{,}229.4$  & $24{,}934$ & $9.39{\times}10^{-10}$ & $1.9{\times}10^{-10}$ & 3.33 & 3.00 \\
 & Qubrio (DS-V3.2)   & $   188{,}228.8$  & $35{,}763$ & $1.00{\times}10^{-10}$ & $2.1{\times}10^{-11}$ & 2.50 & 3.00 \\
 & Qubrio (GPT-5.4)   & $    71{,}201.3$  & $13{,}528$ & $1.13{\times}10^{-8}$  & $2.3{\times}10^{-9}$  & 1.00 & 2.00 \\
\midrule
\multirow{5}{*}{QAOA-60}
 & Enola              & $ 73{,}590.2$ & 0       & $0.0106$ & 0       & N/A & N/A \\
 & PowerMove          & $ 13{,}647.6$ & 0       & $0.1486$ & 0       & N/A & N/A \\
 & Qubrio (Qwen-3.5)  & $ 25{,}320.8$ & $5{,}791$ & $0.0979$ & $0.024$ & 2.66 & 3.00 \\
 & Qubrio (DS-V3.2)   & $ 14{,}475.4$ & $2{,}967$ & $0.1610$ & $0.040$ & 1.50 & 3.17 \\
 & Qubrio (GPT-5.4)   & $ 11{,}847.6$ & $2{,}129$ & $0.1752$ & $0.041$ & 1.33 & 3.00 \\
\midrule
\multirow{5}{*}{GHZ-78}
 & Enola              & $ 21{,}754.9$ & 0       & $0.1720$ & 0       & N/A & N/A \\
 & PowerMove          & $  6{,}424.8$ & 0       & $0.4507$ & 0       & N/A & N/A \\
 & Qubrio (Qwen-3.5)  & $  3{,}500.3$ & $54$    & $0.4798$ & $0.009$ & 2.50 & 3.66 \\
 & Qubrio (DS-V3.2)   & $  1{,}066.1$ & $20$    & $0.5882$ & $0.011$ & 1.66 & 3.00 \\
 & Qubrio (GPT-5.4)   & $    421.9$   & $6$     & $0.6176$ & $0.011$ & 1.00 & 3.00 \\
\midrule
\multirow{5}{*}{Ising-66}
 & Enola              & $ 77{,}380.5$ & 0       & $0.0100$ & 0       & N/A & N/A \\
 & PowerMove          & $  9{,}610.0$ & 0       & $0.2998$ & 0       & N/A & N/A \\
 & Qubrio (Qwen-3.5)  & $  1{,}655.1$ & $921$   & $0.4629$ & $0.046$ & 7.00 & 3.33 \\
 & Qubrio (DS-V3.2)   & $  4{,}060.5$ & $2{,}010$ & $0.3957$ & $0.040$ & 4.00 & 3.00 \\
 & Qubrio (GPT-5.4)   & $    548.4$   & $251$    & $0.4756$ & $0.041$ & 1.00 & 3.00 \\
\bottomrule

\end{tabular}}
\vspace{-20pt}
\end{table*}

Table~\ref{tab:main_full} reports the hardware execution time $T_\text{exe}$, end-to-end circuit fidelity $F$, and average LLM retry counts triggered by the Placement and Routing Verifiers. We evaluate Qubrio across LLM backbones and compare it against two deterministic heuristic baselines, Enola and PowerMove. The results are obtained from $N=6$ independent runs for each benchmark and reported as mean values.

Overall, Qubrio achieves favorable execution time and fidelity compared with the heuristic baselines, with particularly notable runtime improvements on regularly structured circuits. These circuits offer greater opportunities for convoy-style shuttling, allowing compatible displacement vectors to be coordinated and reducing execution overhead. The verifier retry counts also remain within the per-stage budget of 10 across all benchmarks. All runs successfully converge to schedules that pass deterministic verification, demonstrating the effectiveness of the proposed framework.

For Ising-66, we observe relatively large runtime variability across independent runs. This behavior is primarily associated with its repeated local interaction patterns, where each qubit interacts with multiple nearby qubits in succession. Routing decisions that minimize immediate movement costs may lead to unfavorable atom placements for subsequent interactions, requiring additional corrective movements. Consequently, different early routing choices can produce substantially different shuttle schedules and execution times. This suggests that the observed variability stems from the sensitivity of routing decisions to future interactions, rather than verifier instability.

\section{Adaptability Comparison with PowerMove}
\label{app:adapt}
\begin{figure*}[tp]
    \centering
    \includegraphics[width=\linewidth, trim=0in 0in 0in 0in, clip]{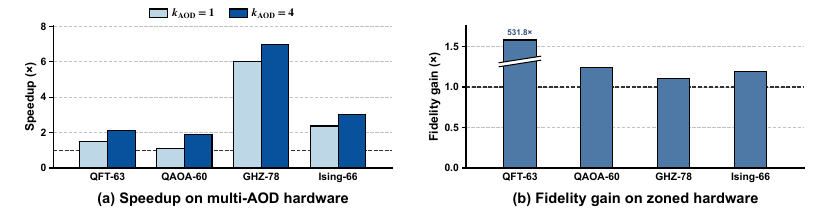}
        \vspace{-10pt}
    \caption{\textbf{Adaptability Comparison with PowerMove.}}
    \vspace{-10pt}
    \label{fig:adapt_powermove}
\end{figure*}

We further evaluate Qubrio's adaptability under multi-AOD and multi-zone hardware configurations, as shown in Fig.~\ref{fig:adapt_powermove}(a) and~(b), respectively. We compare Qubrio against PowerMove using its official adaptations for both scenarios. These adaptations involve algorithmic modifications to nine core modules and changes to more than 90\% of the source files. In contrast, adapting Qubrio requires only tens of lines of declarative hardware specifications and constraint updates, without modifying its core algorithms. Across both scenarios, Qubrio consistently outperforms PowerMove, demonstrating its ability to accommodate different hardware configurations with substantially less manual algorithm-design effort.

\section{Comparison with Concurrent LLM-Based Compilation}

We further discuss the differences between Qubrio and the concurrent work SimpleTES \cite{ye2026evaluation}, which represent two complementary approaches to LLM-assisted compilation. SimpleTES uses an LLM to generate a placement program, while its routing and evaluation components remain hand-written and fixed. Once generated, the resulting compiler processes circuits without further LLM involvement. This can be viewed as a form of distillation that incorporates LLM-generated strategies into a reusable compiler, avoiding repeated LLM inference. In contrast, Qubrio directly integrates specialized LLM agents into the compilation process, enabling more flexible, circuit-specific decisions. 

\begin{figure*}[h]
    \centering
    \includegraphics[width=\linewidth, trim=0in 0in 0in 0in, clip]{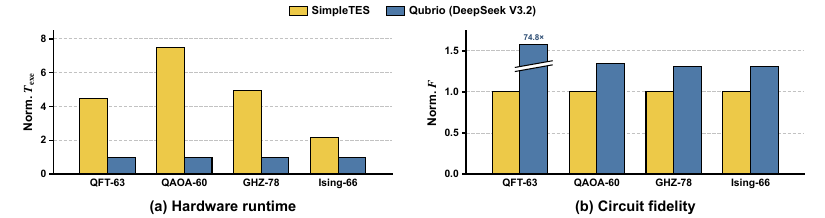}
        \vspace{-10pt}
    \caption{\textbf{Comparison with SimpleTES \cite{ye2026evaluation}.}}
    \label{fig:simpleTES}
\end{figure*}

We also empirically compare the two approaches using the zoned architecture evaluation setup provided by SimpleTES, with identical benchmark circuits, hardware layouts, storage-site spacing, hardware parameters, simulation models, and metric accounting. As shown in Fig.~\ref{fig:simpleTES}, Qubrio consistently achieves lower hardware execution time and higher circuit fidelity across all benchmarks. These results are consistent with the distinction between the two approaches: while distilling LLM-generated strategies into fixed compiler components reduces inference overhead, retaining LLM reasoning during compilation allows more flexible optimization tailored to individual circuits.

\section{Parallel Compilation Strategy}\label{app:parallel}

Qubrio adopts a two-phase parallel compilation strategy to exploit concurrency while respecting the dependency between placement and routing. In the first phase, circuit stages are partitioned into groups, which are processed in parallel. Within each group, stage placements are generated sequentially, using the placement from the preceding stage as a warm start to improve coordination. We empirically set the group size to six, balancing parallelism and placement quality, with less than 10\% performance degradation compared with fully sequential processing. In the second phase, once all stage placements are available, routing tasks for different stages are executed independently and in parallel. Consequently, the overall compilation latency is determined by the slowest placement group followed by the slowest routing task, rather than growing linearly with the number of circuit stages.

\section{Hardware Simulator Configuration and Evaluation Protocol}
\label{app:simulator}

\vspace{0.3em}
\noindent \textbf{Hardware setting.}
Unless otherwise specified, we evaluate each $n$-qubit program on a neutral-atom architecture with one AOD array and an SLM grid of size
$2\times\lceil\sqrt{n}\rceil\times\lceil\sqrt{n}\rceil$.
The layout consists of $\lceil\sqrt{n}\rceil\times\lceil\sqrt{n}\rceil$ gate sites, with two SLM traps hosted at each site. Adjacent gate sites are separated by an inter-site pitch of $15\,\mu\mathrm{m}$, while the two traps within the same gate site are separated by $1.5\,\mu\mathrm{m}$. Consequently, the physical footprint scales as
$16.5\lceil\sqrt{n}\rceil \times 15\lceil\sqrt{n}\rceil\,\mu\mathrm{m}^2$,
where the horizontal pitch $16.5\,\mu\mathrm{m}$ accounts for both the inter-site spacing and the intra-site trap separation.

\vspace{0.3em}
\noindent \textbf{Metrics.}
We evaluate compiler performance using two primary metrics: hardware runtime ($T_{exe}$) and circuit fidelity ($f_{out}$).

\vspace{0.2em}
\noindent \emph{Hardware runtime.}
$T_{\mathrm{exe}}$ denotes the end-to-end hardware runtime of the compiled circuit on neutral-atom hardware. It is computed by summing the durations of all scheduled single-qubit gates, two-qubit gates, atom movements, and SLM--AOD transfers:
\[
T_{\mathrm{exe}}
=
T_{\mathrm{1Q}}
+
T_{\mathrm{CZ}}
+
T_{\mathrm{move}}
+
T_{\mathrm{transfer}}.
\]
We compute $T_{\mathrm{exe}}$ by summing the durations of all scheduled gate, movement, and transfer segments in the compiler-generated execution schedule.

\vspace{0.2em}
\noindent \emph{Circuit fidelity.}
Circuit fidelity captures the end-to-end reliability of the compiled schedule. Following PowerMove~\cite{ruan2025powermove}, we factor the output fidelity into gate, spectator, transfer, and decoherence terms:
\[
f_{\mathrm{out}}
=
f_1^{g_1}
\cdot
f_2^{g_2}
\cdot
f_{\mathrm{spec}}^{\sum_{i=1}^{S} n_i}
\cdot
f_{\mathrm{trans}}^{N_{\mathrm{trans}}}
\cdot
\prod_{q}
\left(
1-\frac{T_q}{T_2}
\right).
\]
Here, $f_1$ and $f_2$ denote the fidelities of single-qubit gates and CZ gates, respectively, and $g_1$ and $g_2$ are their corresponding operation counts. The spectator term $f_{\mathrm{spec}}^{\sum_i n_i}$ accounts for errors on non-interacting qubits exposed during Rydberg pulses, where $S$ is the number of Rydberg pulses and $n_i$ is the number of spectator qubits during the $i$-th pulse. The transfer term $f_{\mathrm{trans}}^{N_{\mathrm{trans}}}$ models the cumulative fidelity loss from $N_{\mathrm{trans}}$ SLM--AOD transfers. Finally, the product over qubits models decoherence, where $T_q$ is the total elapsed time experienced by qubit $q$ and $T_2$ is the coherence time. The hardware parameters used in our simulator are summarized in Table~\ref{tab:NA_hardware_parameter}.

\begin{table}[htbp]
\centering
\small
\renewcommand{\arraystretch}{1.08}
\begin{tabular}{|c||c|c|c|c|c|}
\hline
 & $R_z$ Gate & $R_{x/y}$ Gate & CZ Gate & Spectator & Transfer \\
\hline
Fidelity & $99.9\%$ & $99.9\%$ & $99.5\%$ & $99.75\%$ & $99.9\%$ \\
\hline
Duration & $1\,\mu\mathrm{s}$ & $3\,\mu\mathrm{s}$ & $270\,\mathrm{ns}$ & $270\,\mathrm{ns}$ & $15\,\mu\mathrm{s}$ \\
\hline
\hline
\multicolumn{1}{|c||}{} & \multicolumn{5}{c|}{Qubit Movement} \\
\hline
\multicolumn{1}{|c||}{Fidelity}
& \multicolumn{5}{c|}{$\approx 100\%$ when $a < 2750\,\mathrm{m}\cdot\mathrm{s}^{-2}$} \\
\hline
\multicolumn{1}{|c||}{Duration}
& \multicolumn{5}{c|}{e.g., $100\,\mu\mathrm{s}$ for $27.5\,\mu\mathrm{m}$; $200\,\mu\mathrm{s}$ for $110\,\mu\mathrm{m}$} \\
\hline
\multicolumn{1}{|c||}{Coherence Time}
& \multicolumn{5}{c|}{$1.5\,\mathrm{s}$} \\
\hline
\end{tabular}
\vspace{3pt}
\caption{Neutral-atom hardware parameters used in the simulator~\cite{bluvstein2022quantum, bluvstein2024logical, bochen2024compilation, ruan2025powermove}.}
\label{tab:NA_hardware_parameter}
\vspace{-10pt}
\end{table}

\vspace{0.3em}
\noindent \textbf{Baseline configuration.}
All baselines are evaluated under the same hardware layout, timing model, and fidelity model described above. We use the simulator and default hardware parameters from PowerMove~\cite{ruan2025powermove} and Enola~\cite{bochen2024compilation}, and apply the same post-compilation evaluation script to all generated schedules. For Qubrio, LLM-generated schedules are accepted only after passing deterministic placement and routing verification; invalid schedules are recorded as '-'.

\newpage

\section*{NeurIPS Paper Checklist}

\begin{enumerate}

\item {\bf Claims}
    \item[] Question: Do the main claims made in the abstract and introduction accurately reflect the paper's contributions and scope?
    \item[] Answer: \answerYes{} 
    \item[] Justification: The abstract and introduction clearly state the paper's main contributions and scope, including the agentic compiler design, convoy-style shuttling discovery, and empirical improvements over PowerMove. These claims are supported by the framework description in Sections \ref{sec:method}--\ref{sec:case_study} and the experimental results and ablations in Section \ref{sec:evaluation}.

\item {\bf Limitations}
    \item[] Question: Does the paper discuss the limitations of the work performed by the authors?
    \item[] Answer: \answerYes{} 
    \item[] Justification: The paper discusses the limitations in the Section \ref{sec:conclusion}, noting that current evaluations are limited to NA hardware and that empirically evaluating the framework on other quantum platforms remain future work.

\item {\bf Theory assumptions and proofs}
    \item[] Question: For each theoretical result, does the paper provide the full set of assumptions and a complete (and correct) proof?
    \item[] Answer: \answerNA{} 
    \item[] Justification: The paper does not include formal theoretical results, theorems, or proofs.

    \item {\bf Experimental result reproducibility}
    \item[] Question: Does the paper fully disclose all the information needed to reproduce the main experimental results of the paper to the extent that it affects the main claims and/or conclusions of the paper (regardless of whether the code and data are provided or not)?
    \item[] Answer: \answerYes{} 
    \item[] Justification: The paper describes the workflow in detail. It also provides the code link, and the appendix includes the prompts used by Qubrio.

\item {\bf Open access to data and code}
    \item[] Question: Does the paper provide open access to the data and code, with sufficient instructions to faithfully reproduce the main experimental results, as described in supplemental material?
    \item[] Answer: \answerYes{} 
    \item[] Justification: The code is publicly available at \url{https://github.com/crosaa/Qubrio}, with instructions to reproduce the main results.

\item {\bf Experimental setting/details}
    \item[] Question: Does the paper specify all the training and test details (e.g., data splits, hyperparameters, how they were chosen, type of optimizer) necessary to understand the results?
    \item[] Answer: \answerYes{} 
    \item[] Justification: The paper specifies the experimental setting in Section \ref{subsec:setup}, including the evaluated LLM backbones, benchmark circuits, baseline compilers, hardware configurations, and evaluation metrics. Additional details such as parameters, retry budgets, variance, and simulator settings are provided in the appendices.

\item {\bf Experiment statistical significance}
    \item[] Question: Does the paper report error bars suitably and correctly defined or other appropriate information about the statistical significance of the experiments?
    \item[] Answer: \answerYes{} 
    \item[] Justification: The paper reports run-to-run variance for the evaluated LLM backbones in Appendix \ref{app:raw_data}, providing appropriate information about experimental variability. The main results focus on average hardware runtime and fidelity comparisons, with variance details included in the appendix.

\item {\bf Experiments compute resources}
    \item[] Question: For each experiment, does the paper provide sufficient information on the computer resources (type of compute workers, memory, time of execution) needed to reproduce the experiments?
    \item[] Answer: \answerYes{} 
    \item[] Justification: Details regarding the compute resources used for our experiments are provided in Appendix~\ref{app:compute}.
    
\item {\bf Code of ethics}
    \item[] Question: Does the research conducted in the paper conform, in every respect, with the NeurIPS Code of Ethics \url{https://neurips.cc/public/EthicsGuidelines}?
    \item[] Answer: \answerYes{} 
    \item[] Justification:The research conforms to the NeurIPS Code of Ethics. It does not involve human subjects, private data, surveillance data, or unsafe model release.

\item {\bf Broader impacts}
    \item[] Question: Does the paper discuss both potential positive societal impacts and negative societal impacts of the work performed?
    \item[] Answer: \answerYes{} 
    \item[] Justification: The paper discusses broader impacts, highlighting the benefits of reducing manual compiler redesign and accelerating quantum computing research. Conversely, it also addresses potential risks, including the dual-use nature of quantum applications and the reliance on commercial LLM APIs.
    
\item {\bf Safeguards}
    \item[] Question: Does the paper describe safeguards that have been put in place for responsible release of data or models that have a high risk for misuse (e.g., pre-trained language models, image generators, or scraped datasets)?
    \item[] Answer: \answerNA{} 
    \item[] Justification: The paper does not release high-risk models or scraped datasets.

\item {\bf Licenses for existing assets}
    \item[] Question: Are the creators or original owners of assets (e.g., code, data, models), used in the paper, properly credited and are the license and terms of use explicitly mentioned and properly respected?
    \item[] Answer: \answerYes{} 
    \item[] Justification: The paper credits the existing tools and models used by Qubrio. It also follows their licenses and terms of use.

\item {\bf New assets}
    \item[] Question: Are new assets introduced in the paper well documented and is the documentation provided alongside the assets?
    \item[] Answer: \answerYes{} 
    \item[] Justification: The paper releases new code, prompts, and simulation data. These assets are documented with usage instructions and reproduction details.

\item {\bf Crowdsourcing and research with human subjects}
    \item[] Question: For crowdsourcing experiments and research with human subjects, does the paper include the full text of instructions given to participants and screenshots, if applicable, as well as details about compensation (if any)? 
    \item[] Answer: \answerNA{} 
    \item[] Justification: The paper does not involve crowdsourcing. It also does not involve research with human subjects.

\item {\bf Institutional review board (IRB) approvals or equivalent for research with human subjects}
    \item[] Question: Does the paper describe potential risks incurred by study participants, whether such risks were disclosed to the subjects, and whether Institutional Review Board (IRB) approvals (or an equivalent approval/review based on the requirements of your country or institution) were obtained?
    \item[] Answer: \answerNA{} 
    \item[] Justification: The paper does not involve human subjects. Therefore, IRB approval is not applicable.

\item {\bf Declaration of LLM usage}
    \item[] Question: Does the paper describe the usage of LLMs if it is an important, original, or non-standard component of the core methods in this research? Note that if the LLM is used only for writing, editing, or formatting purposes and does \emph{not} impact the core methodology, scientific rigor, or originality of the research, declaration is not required.
    \item[] Answer: \answerYes{} 
    \item[] Justification:  LLM usage is a core component of the proposed method and is described throughout the paper. The paper also explains that LLM outputs are used as proposals and are accepted only after deterministic verification and simulator-based evaluation.

\end{enumerate}
\end{document}